\documentclass[10pt,aip,apl,twocolumn,superscriptaddress,reprint]{revtex4-1}
\usepackage{graphicx}
\usepackage{textcomp}
\usepackage{color}
\usepackage{cmap}

\begin{document}
\title{Inhomogeneity of donor doping in SrTiO$_3$ substrates studied by fluorescence-lifetime imaging microscopy}

\author{C. Rodenb\"ucher}
\email[]{c.rodenbuecher@fz-juelich.de}
\affiliation{Forschungszentrum J\"ulich GmbH, Peter Gr\"unberg Institut (PGI-7) 52425 J\"ulich, Germany}
\affiliation{Forschungszentrum J\"ulich GmbH, JARA - Fundamentals of Future Information Technologies, 52425 J\"ulich, Germany}
\author{T. Gensch}
\thanks{C. Rodenb\"ucher and T. Gensch contributed equally to this work.\\Copyright 2013 American Institute of Physics. This article may be downloaded for personal use only. Any other use requires prior permission of the author and the American Institute of Physics. }
\affiliation{Forschungszentrum J\"ulich GmbH, Institute of Complex Systems (ICS-4), 52425 J\"ulich, Germany}
\author{W. Speier}
\affiliation{Forschungszentrum J\"ulich GmbH, JARA - Fundamentals of Future Information Technologies, 52425 J\"ulich, Germany}
\author{U. Breuer}
\affiliation{Forschungszentrum J\"ulich GmbH, Zentralinstitut f\"ur Engineering, Elektronik und Analytik (ZEA-3), 52425 J\"ulich, Germany}
\author{M. Pilch}
\affiliation{Forschungszentrum J\"ulich GmbH, Peter Gr\"unberg Institut (PGI-7) 52425 J\"ulich, Germany}
\affiliation{Forschungszentrum J\"ulich GmbH, JARA - Fundamentals of Future Information Technologies, 52425 J\"ulich, Germany}
\affiliation{University of Silesia, A. Che\l kowski Institute of Physics, 40-007 Katowice, Poland}
\author{H. Hardtdegen}
\affiliation{Forschungszentrum J\"ulich GmbH, Peter Gr\"unberg Institut (PGI-9) 52425 J\"ulich, Germany}
\affiliation{Forschungszentrum J\"ulich GmbH, JARA - Fundamentals of Future Information Technologies, 52425 J\"ulich, Germany}
\author{M. Mikulics}
\affiliation{Forschungszentrum J\"ulich GmbH, Peter Gr\"unberg Institut (PGI-9) 52425 J\"ulich, Germany}
\affiliation{Forschungszentrum J\"ulich GmbH, JARA - Fundamentals of Future Information Technologies, 52425 J\"ulich, Germany}
\author{E. Zych}
\affiliation{University of Wroc\l aw, Faculty of Chemistry, 50-383 Wroc\l aw, Poland}
\author{R. Waser}
\affiliation{Forschungszentrum J\"ulich GmbH, Peter Gr\"unberg Institut (PGI-7) 52425 J\"ulich, Germany}
\affiliation{Forschungszentrum J\"ulich GmbH, JARA - Fundamentals of Future Information Technologies, 52425 J\"ulich, Germany}
\affiliation{RWTH Aachen, Institut f\"ur Werkstoffe der Elektrotechnik 2, 52056 Aachen, Germany}
\author{K. Szot}
\affiliation{Forschungszentrum J\"ulich GmbH, Peter Gr\"unberg Institut (PGI-7) 52425 J\"ulich, Germany}
\affiliation{Forschungszentrum J\"ulich GmbH, JARA - Fundamentals of Future Information Technologies, 52425 J\"ulich, Germany}
\affiliation{University of Silesia, A. Che\l kowski Institute of Physics, 40-007 Katowice, Poland}

\begin{abstract}
Fluorescence-lifetime imaging microscopy (FLIM) was applied to investigate the donor distribution in SrTiO$_3$ single crystals. On the surfaces of Nb- and La-doped SrTiO$_3$, structures with different fluorescence intensities and lifetimes were found that could be related to different concentrations of Ti$^{3+}$. Furthermore, the inhomogeneous distribution of donors caused a non-uniform conductivity of the surface, which complicates the production of potential electronic devices by the deposition of oxide thin films on top of doped single crystals. Hence, we propose FLIM as a convenient technique (length scale: 1  \textmu m) for characterizing the quality of doped oxide surfaces, which could help to identify appropriate substrate materials. The following article appeared in Appl. Phys. Lett. 103, 162904 and may be found at http://link.aip.org/link/?apl/103/162904.
\end{abstract}

\maketitle
Perovskite oxides are promising materials for future electronics due to their ability to change their properties under external field gradients. They can induce resistive switching based on an insulator-to-metal transition, which allows the design of redox-based random access
memories (ReRAM).\cite{Waser2007,Waser2009} Bombarding perovskite surfaces with ions can lead to the formation of conducting layers with metallic behavior, which could potentially be used as transparent conductors.\cite{Reagor2005,Rodenbuecher2013} A further method of inducing an insulator-to-metal transition is donor doping, which offers the possibility of growing metallic oxide crystals that can serve as substrates for the growth of functional thin films. For this purpose, a homogeneous distribution of the metallicity related to the donors is needed but meeting this demand is very challenging. For example, it has been shown that substrates with nominally the same doping concentration can have different properties due to a non-uniformity of the dopants.\cite{Huang2006} In this paper, we present investigations of the homogeneity of Verneuil-grown SrTiO$_3$ single crystals donor-doped with Nb and La obtained by three complementary techniques: fluorescence-lifetime imaging microscopy (FLIM), secondary ion mass spectrometry (SIMS), and local-conductivity atomic force microscopy (LC-AFM). We show that FLIM is a particularly fast and cost-efficient characterization technique. It is well suited for investigating the dopant distribution in SrTiO$_3$ since the aliovalent doping leads to modifications in the electronic structure in relation to a valence change from Ti$^{4+}$ to Ti$^{3+}$, which - at a sufficiently high level of critical carrier concentrations - leads to an insulator-to-metal transition.\cite{Smyth1985} It was found that this valence change is related to a blue fluorescence (2.9 eV), which was observed for Nb- and La-doped as well as for Ar$^+$ bombarded SrTiO$_3$.\cite{Kan2005,Kan2006,Rubano2009, Kanemitsu2011,Yang2011} Although the details of the fluorescence process are very complex, it is generally assumed that Ti$^{3+}$ energy levels 0.3~eV below the conduction band generated by oxygen vacancies or donors play an important role.\cite{Yang2011} In doped or activated systems the fluorescence typically shows a maximum for a specific dopant concentration.  This occurs as for higher contents of dopants competitive processes of energy relaxation such as non-radiative Auger recombination become continuously more efficient as reported in particular for donor-activated SrTiO$_3$. \cite{Kan2006, Yasuda2008, Yamada2009}    
Hence, blue fluorescence is a very sensitive characterization property for the intermediate state between insulator and metal. By using a fluorescence microscope, which was operated with two-photon excitation of a fs-pulsed light source\cite{Denk1990}, we were able to measure fluorescence intensity and lifetime with a spatial resolution of 1 \textmu m, and to correlate the detected inhomogeneities with the dopant distribution by SIMS.
Furthermore, we investigated the local conductivity by LC-AFM in order to illustrate the consequences of inhomogeneous conductivity for potential applications. 

Six different SrTiO$_3$ single crystals donor doped with Nb (0.1~wt\%, 0.7~wt\%, 5.0~wt\%) and La (0.075~wt\%, 0.75~wt\%, 3.75~wt\%) purchased from Mateck, Crystec and SurfaceNet were investigated. The samples, which were grown by the Verneuil method, were oriented in (100) direction and the surfaces were epi-polished. Local conductivity measurements were obtained by a JEOL atomic force microscope operating in contact mode under UHV conditions equipped with Pt/Ir tips. Fluorescence spectra were measured after excitation with a 325~nm HeCd laser with a power density of 0.1~W/cm$^2$. FLIM was performed on an upright scanning fluorescence microscope that uses a pulsed high repetition rate Ti:Sa-laser (100~fs, 80~MHz) for excitation and a photomultiplier in non-descanned configuration as a detector \cite{Kaneko2002,Potzkei2012}. Fluorescence intensity decays were reconstructed and analyzed at every pixel by time-correlated single photon counting \cite{Becker2010} with suitable hardware and software. Time-of-flight secondary ion mass spectrometry (ToF SIMS) was used to investigate the distribution of donors by spatially resolved measurements.
\begin{figure}[ht!]
\includegraphics[width=85mm]{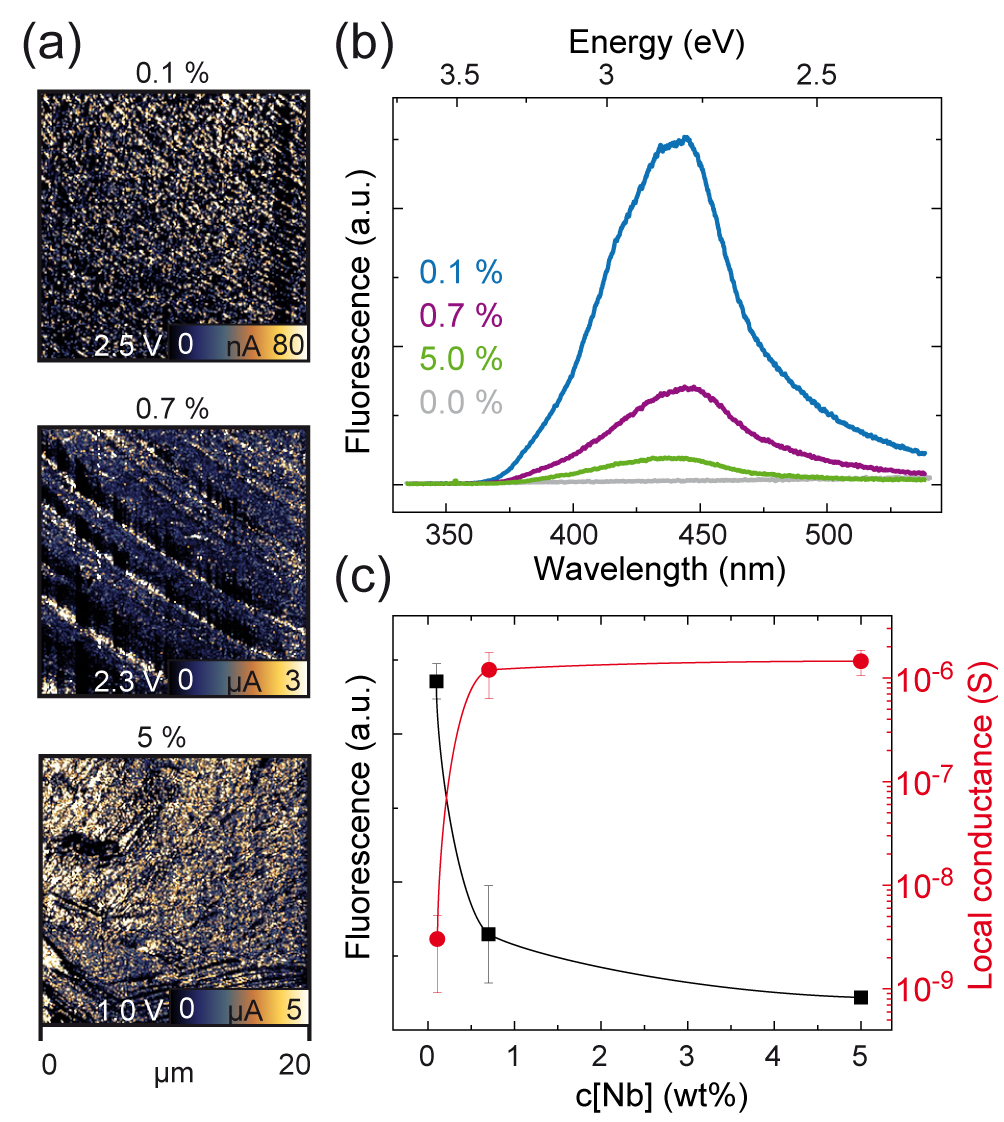}
\caption{Relation between fluorescence and conductivity. (a) LC-AFM current maps. (b) Fluorescence spectra. (c) Fluorescence and conductance as function of donor concentration.}
\label{afm}
\end{figure}

In order to gain an insight into the correlation between donor concentration, local conductivity and fluorescence, a series of Nb-doped crystals were measured (0.1~wt\%, 0.7~wt\%, 5.0~wt\%). Using LC-AFM, the local conductance was recorded on freshly cleaved samples. The current maps (Fig.~\ref{afm}a) show an inhomogeneous distribution of conducting clusters on all three samples, and the extracted average conductance increased with the donor concentration. On the same crystals, fluorescence spectra were recorded (Fig.~\ref{afm}b) revealing a maximum signal at 2.7--2.9~eV as expected with a broad half width indicating the existance of a band of Ti states with reduced valences. While the undoped reference sample did not show fluorescence, a distinct signal was found in the doped crystals with an intensity that decreased with the Nb concentration. This indicates that the fluorescence increases only at very low doping concentrations due to the increasing amount of Ti$^{3+}$ states, while at higher concentrations non-radiative Auger processes suppress the fluorescence due to the high density of doped electrons.\cite{Haruyama1997} The comparison between the donor dependence of local conductance and fluorescence intensity (Fig.~\ref{afm}c) shows a clear anti-correlation of these quantities at higher Nb concentrations, indicating that measuring the spatially resolved fluorescence signal could provide information about the distribution of donors and related conductivity.  
Spatially resolved measurements were performed using two-photon excitation with a wavelength of 740 nm resulting in an excitation energy of 3.35 eV, which is sufficient to bridge the bandgap of 3.2 eV.\cite{Yang2011} 
\begin{figure*}[ht!]
\includegraphics[width=160mm]{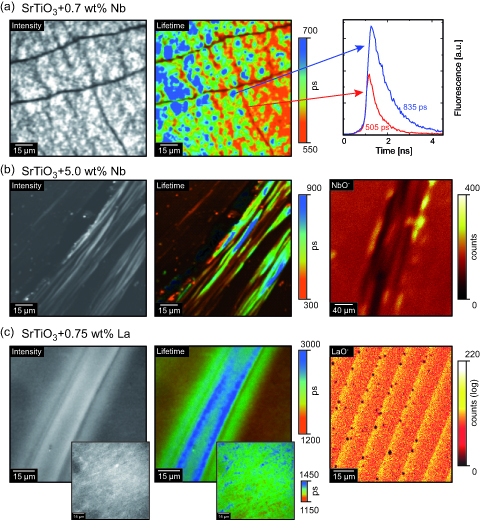}
\caption{Fluorescence intensity and lifetime images measured by FLIM. (a) SrTiO$_3$ + 0.7~wt\% Nb. Inset: Fluorescence intensity decays after fs-pulsed excitation at two positions. (b) SrTiO$_3$ + 5.0~wt\% Nb.  (c) SrTiO$_3$ + 0.75~wt\% La (insets were obtained in the homogeneous part of the sample). The images on the right-hand side show the distributions of NbO and LaO measured by SIMS.}
\label{fluorescence}
\end{figure*}
\begin{figure}[ht!]
\includegraphics[width=85mm]{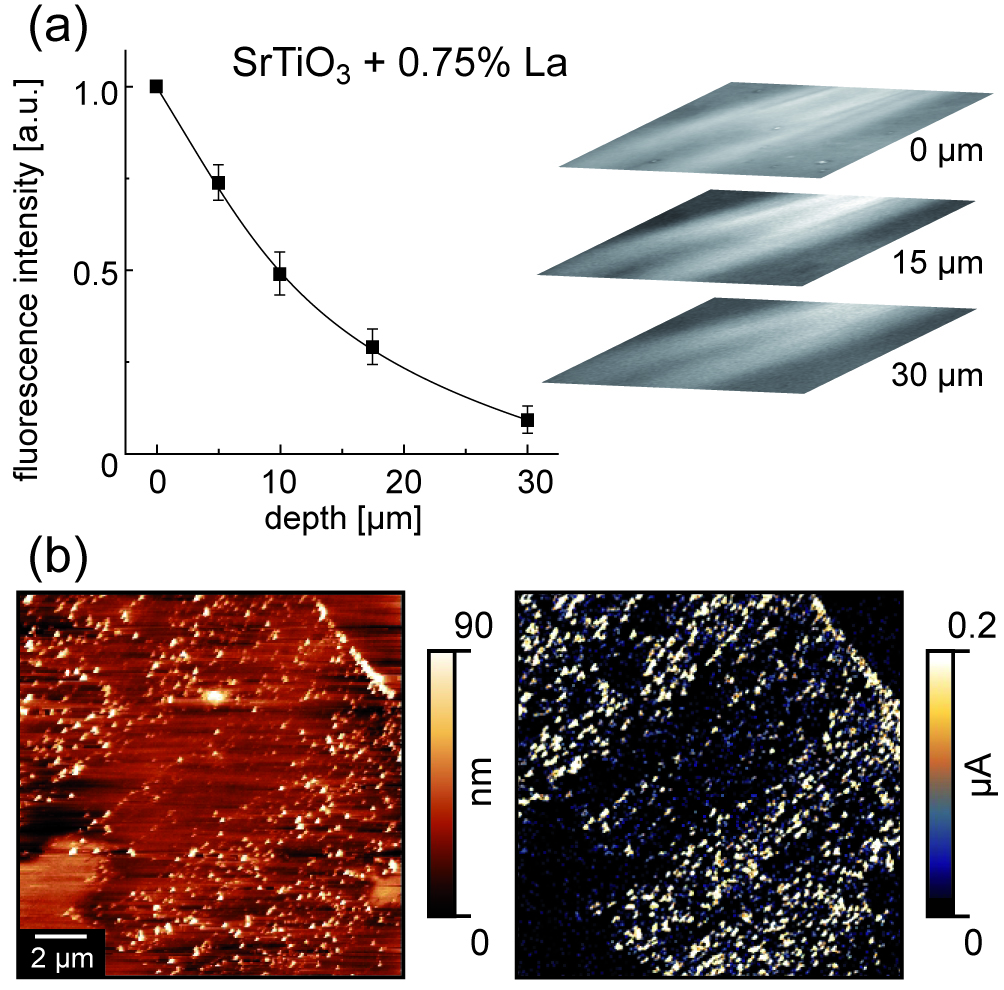}
\caption{(a) Intensity of the fluorescence in SrTiO$_3$ + 0.75~wt\%~La measured at different depths. (b) LC-AFM topography and current maps measured on the surface of SrTiO$_3$:La.} 
\label{depth}
\end{figure}

In Figure \ref{fluorescence}a, the intensity maps and the calculated lifetimes of the fluorescence for Nb:SrTiO$_3$ are presented exemplarily. %While the sample doped with 0.1 \% Nb did not show fluorescence due to reflection on the surface, 
The sample doped with 0.7~\% Nb revealed distinct inhomogeneous cluster-like structures that were found everywhere on the sample. Within these structures, long lines with significantly smaller intensity were detected. The lifetime within these lines was reduced by more than 40\% compared to the rest of the surface, as illustrated in the graph on the right-hand side. On the highest doped sample (5 \% Nb), a different fluorescent structure was found, as shown in Figure \ref{fluorescence}b. In the maps of intensity and lifetime, striped structures with a total thickness of 50-100~\textmu m can be seen. These structures show an additional sub-structure with smaller stripes with a thickness of 3-10~\textmu m. The large differences between the signal inside these stripes also indicated huge differences in the Nb concentration, which was confirmed by spatially resolved secondary ion mass spectrometry. As presented on the right-hand side of Figure \ref{fluorescence}b, in which a high count rate corresponds to a high Nb concentration in the sample, we observed a comparable structure with Nb agglomeration at the rims of the stripe, while the Nb content was decreased inside the stripe. This is in agreement with the fluorescence maps in which a high fluorescence signal was related to a lower donor content according to Fig.~\ref{afm}. Hence, we can conclude that FLIM is a dedicated method for investigating doping distributions with a resolution of 500 nm to 1 \textmu m. 

In order to investigate the behavior of a second donor, we measured the fluorescence of La-doped SrTiO$_3$ with three different concentrations. Fluorescence was found in a similar energy region as in SrTiO$_3$:Nb, which confirms that the modification of the electronic structure does not depend on the specific donor but on the valence change of Ti from 4+ to 3+.  In comparison with Nb-doped SrTiO$_3$, the distribution of the fluorescence is much more homogeneous in all three samples, as shown exemplarily in the insets in Figure \ref{fluorescence}c. Nevertheless, some structures in the fluorescence signal can occasionally be seen, indicating variations in the La concentrations (main images in Figure \ref{fluorescence}c). In both of these images, long stripes can be observed with higher intensities and lifetimes than in the surrounding material. In the SIMS mapping of the La dopant, an inhomogeneous stripe-like distribution was found.  The differences in the shapes of the distribution measured by fluorescence and SIMS may be attributed to the fact that the two measurements were not conducted at exactly the same position, indicating variations of the La distribution on larger scales. However, in both cases an inhomogeneous distribution was detected making it reasonable to assume that the fluorescence signal in SrTiO$_3$:La is also related to the donor concentration.
In order to ascertain whether the inhomogeneities are only a surface effect, we varied the position of the focus of the laser beam perpendicular to the inhomogeneous part of the crystal doped with 0.75 \%~ La. We found that the structure was also present several tens of micrometers below the surface, which is much deeper than typical surface layers in perovskites (Figure \ref{depth}a). In the deeper parts of the sample, the intensity and contrast were reduced, probably due to scattering of the excitation of the luminescence light. However, the inhomogeneities could still be detected. Hence, we concluded that the donor clustering is not only related to surface preparation by cutting or polishing, but that inhomogeneous distribution already evolves during crystal growth. Since the Verneuil growth process takes place far away from the thermodynamic equilibrium depending on the growth parameters, donor clustering can easily evolve.  
Using LC-AFM, we also measured the topography and the conductivity ($U=0.25$~V) of the 0.75~\%La-doped surface.  
As presented exemplarily in Figure \ref{depth}b, an inhomogeneous current pattern was observed verifying that inhomogeneous conductivity related to inhomogeneous donor concentration is not limited to Nb-doped SrTiO$_3$ (Fig.~\ref{afm}a) but rather can be regarded as a general effect of doped SrTiO$_3$.

In summary, we were able to demonstrate the effects of inhomogeneous donor distribution in SrTiO$_3$. Fluorescence microscopy was used to investigate the spatial distribution of the energy states of Ti$^{3+}$ induced by doping with the donors Nb and La. The spatially resolved mapping of fluorescence intensity and lifetime revealed a distinct inhomogeneous distribution of the induced states, which are directly related to donor distribution and the subsequent existence of different conductivity regimes, as verified by SIMS and LC-AFM. This indicates that the transition from insulator to metal induced by donor doping is inhomogeneous in the Verneuil-grown crystals used in our study. This complicates the determination of a precise macroscopic critical density of dopants necessary for metallic conductivity. Furthermore, the existence of such inhomogeneities would cause difficulties when using donor-doped single crystals as metallic substrates for the growth of oxide thin films. If devices were prepared using the crystals investigated, the properties of the devices would depend on their position on the substrate. This emphasizes the need for the investigation on alternative materials and methods for high-quality substrates. Therefore, we propose that a characterization technique employing FLIM could be established in order to estimate the quality and suitability of substrates. In order to further improve this technique, particularly with respect to practical applications, careful calibration measurements using samples (e.g. thin films) of well-known doping concentrations could be performed as they would allow fluorescence intensity to be quantitatively correlated to the doping concentration.

%\begin{acknowledgments}
This work was supported in part by the Deutsche Forschungsgemeinschaft (SFB 917).

%\end{acknowledgments}

% Create the reference section using BibTeX:

%\clearpage

\end{document}